# The inner screen model of consciousness: applying the free energy principle directly to the study of conscious experience

Maxwell J. D. Ramstead[*,1,2,A], Mahault Albarracin[1,3,A], Alex Kiefer[1,4], Brennan Klein[1,5], Chris Fields[6], Karl Friston[1,2], and Adam Safron[7,8]

[1]VERSES Research Lab, Los Angeles, CA 90016, USA
[2]Wellcome Centre for Human Neuroimaging, University College London, London WC1N 3AR, UK
[3]Département d'informatique, Université du Québec à Montréal, 201, Avenue du Président-Kennedy, Montréal, H2X 3Y7
[4]Monash University, Wellington Rd, Clayton VIC 3800, Australia
[5]Network Science Institute, Northeastern University, Boston, Massachusetts, USA
[6]Allen Discovery Center, Tufts University, Medford, MA, USA
[7]Department of Psychiatry & Behavioral Sciences, Johns Hopkins University School of Medicine, Baltimore, MD, USA
[8]Institute for Advanced Consciousness Studies, Santa Monica, CA, USA
[A]These authors contributed equally.

21st April 2023

## Abstract

This paper presents a model of consciousness that follows directly from the free-energy principle (FEP). We first rehearse the classical and quantum formulations of the FEP. In particular, we consider the "inner screen hypothesis" that follows from the quantum information theoretic version of the FEP. We then review applications of the FEP to the known sparse (nested and hierarchical) neuro-anatomy of the brain. We focus on the holographic structure of the brain, and how this structure supports (overt and covert) action.

## Contents

[*]maxwell.ramstead@verses.ai

# Additional information

## Acknowledgements

The authors thank Philippe Blouin, Guillaume Dumas, Jonas Mago, David Rudrauf, Grégoire Sergeant, Lars Sandved Smith, Anil Seth, Toby St Clere Smithe, Jeff Yoshimi, Robert Worden, the members of the CompPhen Fridays discussion group, and the other members of the VERSES Research Lab for valuable comments on early versions of this work, and for useful discussions that shaped the contents of the paper. Special thanks are due to Jakob Hohwy, Mark Solms, Wanja Wiese, and Ken Williford.

## Funding statement

The authors are grateful to VERSES for supporting the open access publication of this paper. BK acknowledges the support of a grant from the John Templeton Foundation (61780). KF is supported by funding for the Wellcome Centre for Human Neuroimaging (Ref: 205103/Z/16/Z) and a Canada-UK Artificial Intelligence Initiative (Ref: ES/T01279X/1). The opinions expressed in this publication are those of the author(s) and do not necessarily reflect the views of the John Templeton Foundation.

# 1 Introduction

It is a truism that, since the late 1990s, the field of consciousness studies has undergone a revival. Since its resurgence, several theories and models of consciousness—both conceptual and formal—have been proposed. Theories of consciousness vary widely in their goals, from addressing the "hard problem" (Chalmers, 1996) of why experience exists at all, to modelling how particular contents of experience are implemented by particular neural structures. They also vary widely in scope, and hence in the variety of "consciousness" that is of interest; ranging from theories of basal awareness applicable to all organisms and perhaps beyond, to theories of specifically human-like consciousness. These latter typically take as explananda not just sensory awareness, but also such phenomena as interoception, the sense of being a self, the sense of being the origin point of experience (or subjectivity), and the ability to make sense of the minds of other agents (aka "theory of mind"). Several of these latter theories share core features, and overlap in terms of their explanatory targets as well as their explanatory posits and assumptions. Indeed, at the time of writing, dozens of theories of consciousness have emerged in computational neuroscience alone: for a recent review, see (Seth & Bayne, 2022).

The aim of this paper is to present a model of consciousness that follow directly from the variational free energy principle (FEP). At a fundamental level, the FEP is concerned with (state) spaces that can be decomposed into some "thing", "particle", or particular "system" of interest, and the rest of state space (Friston, 2019a). Provided that the interaction across the boundary—that individuates some "thing" from everything else—is sufficiently weak, then the states on the interior of the boundary will be conditionally independent or separable from exterior or external states. Both the particular system of interest and its external environment will, in this case, have well-defined state spaces. The FEP characterizes the behaviour of such system-environment pairs, regardless of whether their state spaces are described in classical (Friston, 2019a; Ramstead et al., 2023) or quantum (Fields et al., 2022) terms. The compass of the FEP thus includes all systems that are intuitively regarded as conscious—including all organisms—but also reaches far beyond them. Our contribution here will be to frame an explicit hypothesis about what additional necessary structure must be in play for a system, whether described in classical or quantum terms, to be conscious.

We approach the question of consciousness by starting from the FEP and using it to make sense of established features of (human) neuroanatomy: namely, the nested and hierarchical structure of the brain that arises from a particular kind of sparse coupling. This sparsity structure can be modelled as a set of hierarchically nested Markov blankets; where the Markov blanket of a given system constitutes the interface between that particular system and everything else (Fields, Glazebrook, & Levin, 2022; Hipólito et al., 2021). Note that the system-environment boundary, and hence the location of the Markov blanket within state space, can change over time, e.g., in cases involving material exchange and turnover between the system and its environment (Ramstead & Friston, 2022; Ramstead et al., 2021). We will highlight the importance of actively configuring and maintaining the sparsity structures that underwrite Markov blankets; and suggest that conscious processing can be identified in terms of its contribution to action. For example, consider overt action that is apparent to an observer such as moving: e.g., palpation and saccadic sampling; or covert actions, such as autonomic reflexes or attention and mental action in the brain; see (Sandved-Smith et al., 2021).

Our core argument can be summarized as follows: 1. Empirically, we know that the brain evinces a sparse (i.e., hierarchical or heterarchical, and nested) structure; 2. Using the FEP, we can model any part of the brain, from the sub-neuronal level to distributed neuronal networks, as having Markov blankets. Given this setup, the *dynamical dependencies* that drive the *flow of information* in the brain can be represented as hierarchically nested Markov blankets. Mathematically, this is equivalent to a set of holographic screens nested within one another, lending the set of nested screens a *holographic structure*; 3. The blankets associated with conscious processing can be picked out from the ones that are enabling—but not constitutive—by examining their role in selecting actions.

A few words about prior art. Ours is not the first attempt to develop an integrative treatment of consciousness premised on the FEP. One notable attempt to integrate a broad spectrum of theories of consciousness is integrated world modelling theory (IWMT) (Safron, 2020a, 2022). IWMT represents an attempt to integrate FEP accounts with two other leading theories of consciousness, namely, the global workspace theory (Baars, 2005) (GWT) and integrated information theory (IIT) (Tononi, 2015; Tononi et al., 2016). A second body of work that has focused on the role of active inference in the selection of "winning hypotheses" corresponding to conscious perceptual contents, as in binocular rivalry paradigms (Hohwy, 2013, 2022; Hohwy, Roepstorff, & Friston, 2008; Parr et al., 2019), has also presented an FEP-theoretic perspective on lines of evidence motivating many distinct theories of consciousness. We see the present work as deeply continuous with this prior art. In this work, we focus squarely on what the FEP itself, applied to anatomical and functional brain architectures, may suggest about the structures that realize consciousness. Furthermore, considering that our holographic perspective is grounded in quantum information theory, the explanatory reach of the ensuing model extends beyond brains to encompass more basal, non-neural systems (Fields, Glazebrook, & Levin, 2021). That is, this work can be considered to represent steps towards identifying necessary conditions for the realization of consciousness in all entities—all particles, in the parlance of the FEP—in which such computational properties might be observed.

The argumentative structure of this paper is as follows. We first rehearse the FEP and its applications to functional neuroanatomy. We then present a minimal model of consciousness based on the FEP. We conclude with discussion of core concepts and predictions that emerge from this treatment.

## 2 Preliminaries: A short introduction to the FEP

We first provide some introductory remarks on the FEP. The model considered in this paper is premised on the FEP. This section starts with the classical (random dynamical system) formulation of the FEP and then turns to its formulation in terms of quantum information theory.

## 2.1 The classical formulation of the FEP

The FEP is a scientific or mathematical principle that, much like the principle of least action or the principle of maximum entropy, can be used to derive the mechanics of coupled random dynamical systems (Friston, 2019a; Ramstead, Sakthivadivel, & Friston, 2022; Ramstead et al., 2023). Just as one can think of the principle of least action as the variational principle that underwrites classical mechanics, one can think of the FEP as the variational principle that describes the way that beliefs evolve over time; namely, Bayesian mechanics (Ramstead et al., 2023; Sakthivadivel, 2022b).

The FEP provides an explanation—from first principles—of why it looks as if anything that exists represents or infers the properties of the environment to which it is coupled (Ramstead et al., 2023). Metaphorically, the FEP is a "map" (literally: a scientific model) of that part of the "territory" (literally: the system being modelled) that "behaves as if it were a map" (literally: that tracks features of the modelled system, as unpacked briefly below). In other words, the FEP provides us with a way to understand the dynamics of any physical thing that exists, as itself modelling the statistical structure of its embedding environment (Ramstead, Sakthivadivel, & Friston, 2022).

Intuitively, for some given thing to be or act as a map of some other thing (considered as a territory), it must be, at least in some minimal sense, *distinguishable* or *separable* from that territory. This intuition is formalised in Bayesian mechanics with the construct of *Markov blankets* (Ramstead et al., 2023; Sakthivadivel, 2022b). The core idea is that systems that conform to the FEP are endowed with a boundary, called a Markov blanket, which separates a "thing" or particle from its environment—but also couples the one to the other (Ramstead, Sakthivadivel, & Friston, 2022). A Markov blanket renders the internal states of some thing conditionally independent from the external states of that which it is not. The Markov blanket can be decomposed into sensory and active states (Palacios et al., 2020). By construction, sensory states are those that influence but are not influenced by internal states; whereas, symmetrically and reciprocally, active states influence but are not influenced by external states.

Heuristically, the FEP says that anything that has a Markov blanket—in the sense of being realised over some nontrivial timescale—must be tracking the statistics of its embedding environment (Da Costa et al., 2021). In more detail, the FEP tells us that if a system exists and has a Markov blanket, then it entails a *generative model* of its world. Mathematically, a generative model is a joint probability density over states that influence each other. In this setting, we can express the dynamics of active and internal states in terms of a gradient flow on a variational free energy that depends upon a generative model of how external states cause sensory states; where the word "cause" is used in the straightforward (dynamical systems) sense that one state causes changes in another state because the first state enters the equations of motion of the second. (Ramstead et al., 2023). Effectively, this allows one to interpret the autonomous (i.e., internal and active) dynamics as a process of inference; namely, changing in a way that minimises variational free energy or, equivalently, maximising the evidence for the generative model. This is sometimes called self evidencing (Hohwy, 2016).

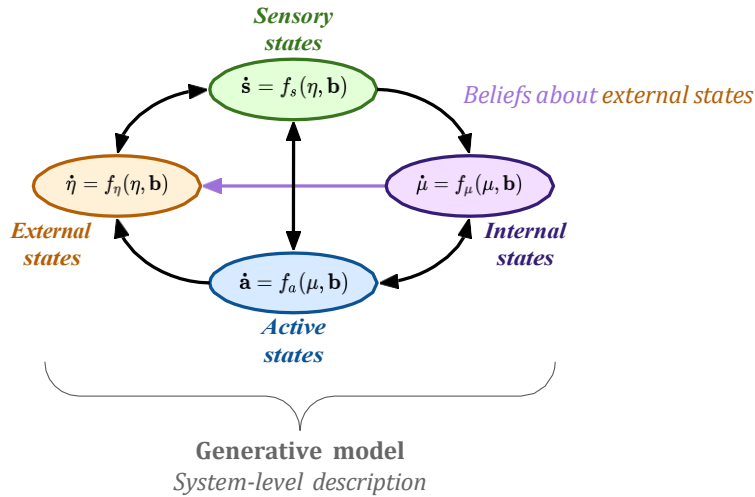

Figure 1: **Illustration of a Markov blanket.** *Internal* and *external* states interact via *sensory* and *active* (i.e., blanket) states, which induce a conditional independence between internal and external states. In virtue of this conditional independence, one can associate internal states with the parameters or sufficient statistics of a conditional (a.k.a., variational) density over external states, given blanket states. In turn, one can then interpret internal (and active) dynamics as a gradient flow on a variational free energy that plays the role of a marginal likelihood or model evidence in statistical inference.

Under the FEP, when a system is partitioned into Markov blankets, the internal states of each Markov blanket can be read as tracking each other (Friston, 2013). In FEP terms, the internal states of a particle will look as if they are updating a probabilistic representation of the causes of its sensory states. (Note that this description can never be verified, because the internal states are inaccessible, provided the Markov blanket remains intact: all that can be observed from the outside are the blanket states.) This representation, parameterised by internal states, is referred to as a variational density over external states.

As a result, particles can be read as tracking their environments through their Markov blankets. Noteworthy to the philosophical reader, this notion of representation as "tracking" differs from more traditional appeals to the kinds of causal-informational dependency that feature in philosophical psycho-semantics (Dretske, 1981; Fodor, 1990; Usher, 2001), in several respects. Broadly speaking, Dretske (1981) and Fodor (1990) (among others) posit atomic relations of lawful covariation between the internal states of agents, on the one hand, and entities or events in the represented world, on the other (what a FEP theorist would call external states). On this account, an internal state represents an external state if, and only if, there exist lawful relations of covariation between both states, such that changes in internal states track changes in external states. In the FEP formulation, the internal states of a system do not *directly* track or covary with the external states of a system: rather, they encode the parameters of a conditional *probability density* over external states—and it is this density which represents external states (Sprevak, 2020). This is a core but crucial difference. That said, this form of mapping relates to "tracking" in the conventional sense, since internal paths of least action encode expectations about external paths (where action is the path integral of variational free energy). Crucially, the factors of the conditional density (e.g., types of objects, outcomes, and events) that "carve up" the world may vary arbitrarily, subject only to the constraint that model evidence is conserved on average (Ramstead et al., 2023; Sakthivadivel, 2022b). This account of representation may thus be regarded as justifying (or dovetailing with) Quine's view that the categories in terms of which we experience the world are akin to theoretical posits, which earn their keep via their predictive and explanatory power with respect to sensation (Quine, 1960). Nonetheless, structural correspondences can be expected to obtain at the level of internal and external states as a whole (Conant & Ashby, 1970; Kiefer & Hohwy, 2018; Ramstead, Kirchhoff, & Friston, 2020), where the KL-divergence between the variational and generative densities has been argued to provide an account of misrepresentation, and thus a form of semantic normativity (Kiefer & Hohwy, 2018; Ramstead, Friston, & Hipólito, 2020).

## 2.2 The quantum information theoretic (scale-free) formulation of the FEP

Quantum information theory provides an interpretation of Markov blankets in light of the holographic principle (Fields et al., 2022). The holographic principle is a fundamental result in physics, which asserts that a given volume of space-time or "bulk" cannot contain more externally-observable or accessible information than can be encoded on its boundary (Almheiri et al., 2021; Bousso, 2002; Fields, Glazebrook, & Marcianò, 2022). The bulk can contain arbitrarily many degrees of freedom that do not contribute to the interaction defined at the boundary, but from the point of view of an outside observer, these hidden degrees of freedom can only, by definition, be inferred from the observable ones (Almheiri et al., 2021). The holographic principle has been applied to a variety of physical systems, including black holes and quantum gravity, and has also been explored in the context of consciousness (Fields, Glazebrook, & Levin, 2021).

In the quantum information theoretic formulation of the FEP, Markov blankets are implemented by holographic screens (Fields et al., 2017). Separability between a system and its embedding environment—and hence identifiability of either—corresponds by definition to absence of entanglement: in this case, the boundary functions as a classical information channel (Fields & Marcianò, 2020). The particle-environment boundary can then be considered a holographic screen (Fields, Glazebrook, & Marcianò, 2022) via which the system and its environment exchange classical information. A holographic screen is, by definition, a boundary in the joint system-environment state space that encodes all of the information that one system can obtain about the other; hence a holographic screen functions as a Markov blanket, in the sense that it assures conditional independence between system and environment states. Conversely, by epistemically or informationally shielding internal from external states (while also coupling them), Markov blankets can—by definition—be under-stood as holographic screens that encode all information that can be transmitted between systems and their environments.[1]

The interaction between any two quantum systems A and B that are jointly isolated—i.e., between any system and its entire environment—can be represented as a cycle in which A "writes" a finite bit string on their mutual boundary *B* that B then "reads," after which B writes a bit string of the same length that A reads in turn (Fields et al., 2022; Fields, Glazebrook, & Marcianò, 2022). The variational free energy for either system can then be defined as the difference between the bit string most recently written and the bit string that is subsequently read, i.e., the string most recently written onto the screen by the environment (Fields et al., 2022). The FEP requires any system to minimize the difference between expectation (the written string) and observation (the string subsequently read). As the limit of this process (in which writes and reads exactly match) corresponds to quantum entanglement, the FEP can be seen to be the classical limit of the principle of unitarity, i.e., the principle of conservation of information, upon which quantum theory is based (Fields et al., 2022).

In a classical setting, separability between a system and environment can be achieved, provided they are only weakly coupled, by separating them in space. The classical formulation of the FEP assumes a space-time background; and thereby, one can define the states of some thing, as against random fluctuations, by appealing to timescale separation: states are things that change slowly enough at some scale to be reliably re-identified as the same states of the same thing (Fields et al., 2022); while other things change so quickly that they average out. Thus, the classical FEP is inherently multi-scale (Heins et al., 2023; Ramstead et al., 2021). The quantum information theoretic formulation does not assume a space-time background; indeed, it is consistent with quantum-gravity models in which spacetime is both emergent from the underlying informational dynamics and system-relative (Fields, Glazebrook, & Marcianò, 2023). It is, therefore, fully scale-free, applying in the same form to all systems; ranging from particle-particle interactions within the standard model, through the scales of molecules, cells, multi-cellular organisms, and biological populations and communities, through to the cosmological scale of quantum fields, black holes, and large-scale structures.

## 2.3 The inner screen hypothesis

In light of the quantum information theoretic formulation of the FEP, one can think of the Markov blanket as a kind of holographic screen, which encodes all the information available to an external observer (Fields, Glazebrook, & Marcianò, 2022). That is, according to the holographic principle, all the information about some thing that is available from the perspective of an external observer must, by definition, be encoded on its surface, i.e., on its Markov blanket (Fields et al., 2017). Recently, it has been proposed that this harnessing of the classical information—that characterizes the coupling between two systems—can be leveraged by those systems themselves, in the service of prediction and control.

---

[1]It should be noted that the term "holography" has been used in two different (but conceptually related) ways the literature, which should be distinguished from each other. On the one hand, the term "holography" refers to a set of techniques allowing for the reconstruction of a three-dimensional image by exploiting the interference patterns of multiple two-dimensional wavefronts. The term "holographic," as used in connection with the holographic principle in physics, retains the essential idea that the information content of a bulk can (from the point of view of an observer) in principle be encoded on a lower-dimensional surface, but without any commitment to encoding via interference patterns. The former sense of holography was in play in previous theories of consciousness, e.g., the holonomic brain theory (Pribram, Yasue, & Jibu, 1991); we focus on the latter sense.

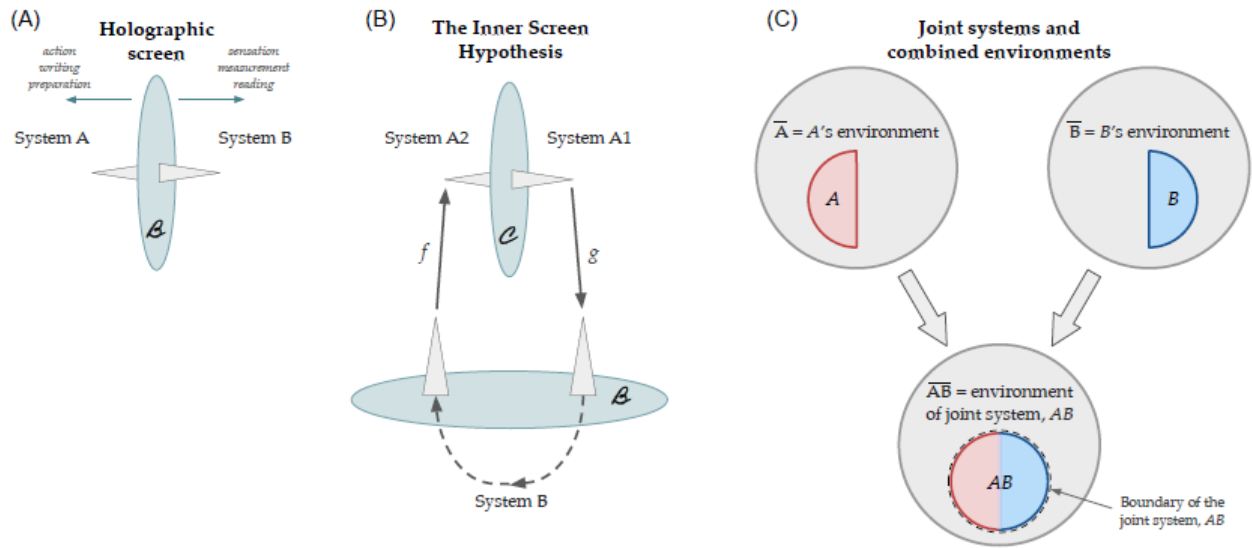

Figure 2: **Holographic screens and their environments.** (A) The boundary $\mathscr{B}$ between any two mutually separable systems A and B is a holographic screen that encodes, as classical data, the eigenvalues of the interaction (i.e., the Hamiltonian operator) that couples A to B (Fields et al., 2022). A's actions on $\mathscr{B}$ are B's perceptions, etc. Triangles represent computational processes (quantum reference frames) implemented by A and B, respectively. (B) A two-component system (A1, A2) interacts with an environment B. The interaction is mediated by an internal boundary between A1 and A2 that serves as a Markov blanket, and hence as an internal classical memory structure. The maps f and g implement "perceptions" of B by A2 and "actions" on B by A1. Graphic adapted and inspired by Fig. 3 of (Fields, Glazebrook, & Levin, 2021). (C) Markov blankets of joint systems comprise the "faces" of the Markov blankets of the component systems that are exposed to their common environment. The environment $\overline{AB}$ of the combined system $AB$ is smaller than either of the environments ($\bar{A}$ or $\bar{B}$) of its components ($A$ or $B$).

A prediction about the requisite architecture for consciousness can be derived directly from the FEP. According to the *inner screen hypothesis* (Fields, Glazebrook, & Levin, 2021), for a system to exhibit any form of consciousness involving even short-term internal (classical) memory requires that the system have an internal Markov blanket separating it into at least two distinguishable components. (Fields et al., 2022; Fields, Glazebrook, & Marcianò, 2022) have suggested that an internal Markov blanket would encode all the classical information that is available internally to a system, and that consciousness itself may be (or entail) classical information encoded on an internal Markov blanket.

Sufficiently sparse coupling between internal subsystems is crucial in admitting an internal Markov blanket structure and hence allowing classical information processing and integration by the system to generate—on this view—conscious experience. Overall, the holographic principle theoretic interpretation of internal Markov blankets provides us with a deep theoretical framework for understanding the relationship between sparsity and consciousness. The inner screen hypothesis is really the claim that a conscious system has to be *composite*: it has to have at least two components that are mutually separable, and which are communicating with each other classically.

The idea of a set internal screens, as well as the holographic principle reading of the Markov blanket, will play a key role in our account. With our model of consciousness, we are proposing to take this hypothesis seriously and to pursue it formally. There is a real sense in which this hypothesis amounts to the positing of an inner *homunculus* (Lycan, 1996)—witness to projections on the internal screen (a "Cartesian theatre"; see (Safron, 2021a)). We suggest, however, that contemporary developments in physics (particularly quantum information theory and the holographic principle), together with the FEP, motivate a qualified Cartesianism that eludes the historical baggage of such positions: in place of a dualism whose terms are an unextended "thinking substance" (mind) and extended matter, we may embrace an (observer-relative) dualism between states internal to a Markov blanket (which constitute the capacity for observation of a given system, without themselves being observable) and those outside it (which constitute the world represented in those observations). There is nonetheless a poetic connection to Descartes's original view, as the idea of the holographic principle was derived from the thermodynamics of black holes, whose internal structures are modelled as singularities (i.e., "substances" which, while possessing "mass", are physically un-extended).

## 2.4 Markov blankets in the brain

Having reviewed the core formulations of the FEP, we turn to its application to the brain's sparse functional anatomy. It has long been known that the brain has a characteristic *sparse* pattern of connectivity at many scales; see, e.g., (Ahmad & Scheinkman, 2019; Bullmore & Sporns, 2012; Felleman & Van Essen, 1991; Gal et al., 2017; Hilgetag, O'Neill, & Young, 2000; Mesulam, 1998; Sporns & Zwi, 2004).

The FEP has been used to make sense of the nested structure and hierarchical networks of the brain. Indeed, the FEP was originally formulated and applied in the context of computational neuroscience, to account for the structure, function, and dynamics of the human brain (Friston, 2005, 2010; Friston, Kilner, & Harrison, 2006). The core idea is that the Markov blankets that individuate a given entity (i.e., separate something from, but couple it to, its embedding environment) arise at many spatial and temporal scales; where the same essential pattern (i.e., the existence of a Markov blanket) repeats across scales (Friston, Fagerholm, et al., 2021). It is important to emphasize here that Markov blankets are never created *de novo*, but rather assemble as the components of the system assemble (Palacios et al., 2020). A simple example is shown in Figure 2: here the Markov blanket of a two-component system comprises the "faces" of the Markov blankets of the component systems that remain exposed to their common environment. This hierarchical assembly process has three immediate consequences: (i) no two components of any multi-component system experience the same environment; (ii) no component experiences the same environment as the entire system; and (iii) the entire system does not experience the same environment as any of its components. Indeed, as evident in Figure 2C, the size of the joint-system environment decreases (and the surface area of its Markov blanket increases) monotonically as components are added to any multi-component system.

Provided that the components of a joint system are not mutually isolated, i.e., non-interacting, they will share a boundary; the components A and B in Figure 2C, for example, share the boundary at which A touches B and vice-versa. Assuming that the A-B interaction is sufficiently weak that A and B remain unentangled and hence evince conditionally independent states, this boundary mediates classical communication between A and B. It thus serves as an "internal" screen or Markov blanket within the composite system comprising A and B. Such internal screens are required for internal (classical) information flow, for internally-stored memory, and for an internal representation of the passage of time (Fields, Glazebrook, & Levin, 2021). Hence, composite systems—with internal screens implementing Markov blankets—are essential in any system that experiences any form of consciousness beyond the most basal, unremembered awareness. All organisms, including unicellular ones, have such compartmentalized internal structures.

Recent work has shown that we can interpret the structure of the brain in terms of nested Markov blankets, and hence compartmentalization, at several scales: starting from the scale of individual neurons, we ascend to cortical microcircuits, cortical layers, brain regions, and whole-brain networks, each with their own Markov blanket (as reviewed in (Hipólito et al., 2021)); or we descend towards dendrites, synapses, and cytoskeletal surfaces, each construed in turn as Markov blanketed structures that assemble into Markov blanketed structures at the scale above, as discussed by (Fields, Glazebrook, & Levin, 2022).

We now rehearse the argument about the nested, Markov blanketed structure of the brain, made in greater detail by (Hipólito et al., 2021) and (Fields, Glazebrook, & Levin, 2022). The presence of a Markov blanket formalises the requirement that internal and external states (or paths) are independent from each other, given blanket states. In other words, internal and external states influence each other, but only vicariously or indirectly, via the blanket. Now, Markov blankets are statistical boundaries: that is, they are composed of the degrees of freedom through which a system and its environment interact. Importantly, these degrees of freedom reflect the dynamics and ensuing conditional independencies of a particular system, and need not correspond to, or map one-to-one onto, the spatial boundaries of such systems, e.g., the surfaces of neuronal cells. In other words, the Markov blanket does not cut internal states off from their embedding world, but instead constitutes the means by which the system is coupled to its world; hence the Markov blanket itself can be viewed as marking out which states are relevant to the self-organisation of the system, statistically speaking.

Individual neurons can be modelled as having their own individual Markov blankets. A simple model of a Markov-blanketed neuron maps the internal states of the neuron to the postsynaptic conductance of the neuron, the internal concentrations of relevant ions within the neuron, and states of the signal-transduction, metabolic, and gene-regulatory pathways of the neuron; sensory states correspond to postsynaptic voltages; active states, in turn, can be mapped to pre-synaptic voltages; and finally, the external states of the system can be mapped onto the ion, neurotransmitter, and other molecular concentrations in the immediate extracellular environment of the neuron, together with the states all other proximate neurons, glia, or other cells.

From there, we can go "up" towards increasing scales or "down" towards decreasing scales. One could examine the construction in the descending direction, from the scale of neurons to the scale of their component processes. Indeed, the neuron is itself an ensemble of several, different relatively autonomous components (cell surfaces, dendrites, synapses, and so on); e.g., (Branco, Clark, & Häusser, 2010; Kiebel & Friston, 2011).

In the other direction of travel, we can ascend from the scale of individual neurons to nested scales of increasingly large ensembles of neurons, which form coherently, such that they too have their own Markov blankets. Canonical microcircuits, for instance, have their own blankets at a superordinate scale (Bastos et al., 2012; Douglas & Martin, 1991; Friston, Parr, & de Vries, 2017).

This account also generalises to different nested connectivity structures in the brain: That is, it is not merely that processing at a given scale of self-organisation has a hierarchical architecture; it is that we find formally identical patterns at subordinate and superordinate scales (Friston, Fagerholm, et al., 2021; Ramstead et al., 2019). In turn, brain regions each have their own Markov blankets, and networks of such regions have their own blankets. Such findings—which have often been described in terms of modularity—can be characterized more precisely by appealing to the idea of factorization (Parr, Sajid, & Friston, 2020). That is, brain regions are specialised, such that they encode distinct (probability) factors that are statistically independent. In other words, physiological separation of brain regions entails that the associated information encoded in those regions is conditionally independent of that encoded in other distinct regions (Friston & Buzsáki, 2016; Parr, Sajid, & Friston, 2020).

## 2.5  Hierarchical predictive coding and Markov blankets

The FEP has been used to make sense of hierarchical neuronal message passage at a given scale. In this context, the internal states of a neuronal system (at any scale) encode expectations about sensory signals, which are passed downward to lower levels of a hierarchy. This is known as hierarchical predictive coding (Friston, Breakspear, & Deco, 2012; Hohwy, 2020; Lee & Mumford, 2003; Mumford, 1992; Rao, 1999).

We can bring a fresh perspective on these formulations by considering the Markov blanket of any level of the hierarchy at some fixed scale. In hierarchical predictive coding, at a given scale, each hierarchical level has its own Markov blanket, and exchanges with other levels (those comprising individual neurons, microcircuits, etc.) according to the dependency structure of the Markov blanket (Nikolova et al., 2021). In this scheme, ascending prediction errors can be read as active states, while descending predictions constitute sensory states, from the point of view of any given level (and *vice versa* from the point of view of a supraor-dinate level). This is illustrated in the right panel of Figure 5. We note that the description of levels in a neuronal processing hierarchy in terms of Markov blankets can be generalized beyond the assumptions of predictive coding architectures. For example, in architectures that do not posit distinct error-encoding neurons; e.g. (Scellier & Bengio, 2016), we may still describe a processing layer as having a Markov blanket, whose active states comprise the set of active states of all the individual neurons in the level with connections to other levels; similarly for sensory states. The key thing to observe here is that there are (as least) as many Markov blankets as there are levels in the hierarchical model. We have illustrated the Markov blanket separating neuronal states from environmental (here, motor system) states—at the lowest level of the hierarchy—and an intermediate Markov blanket.

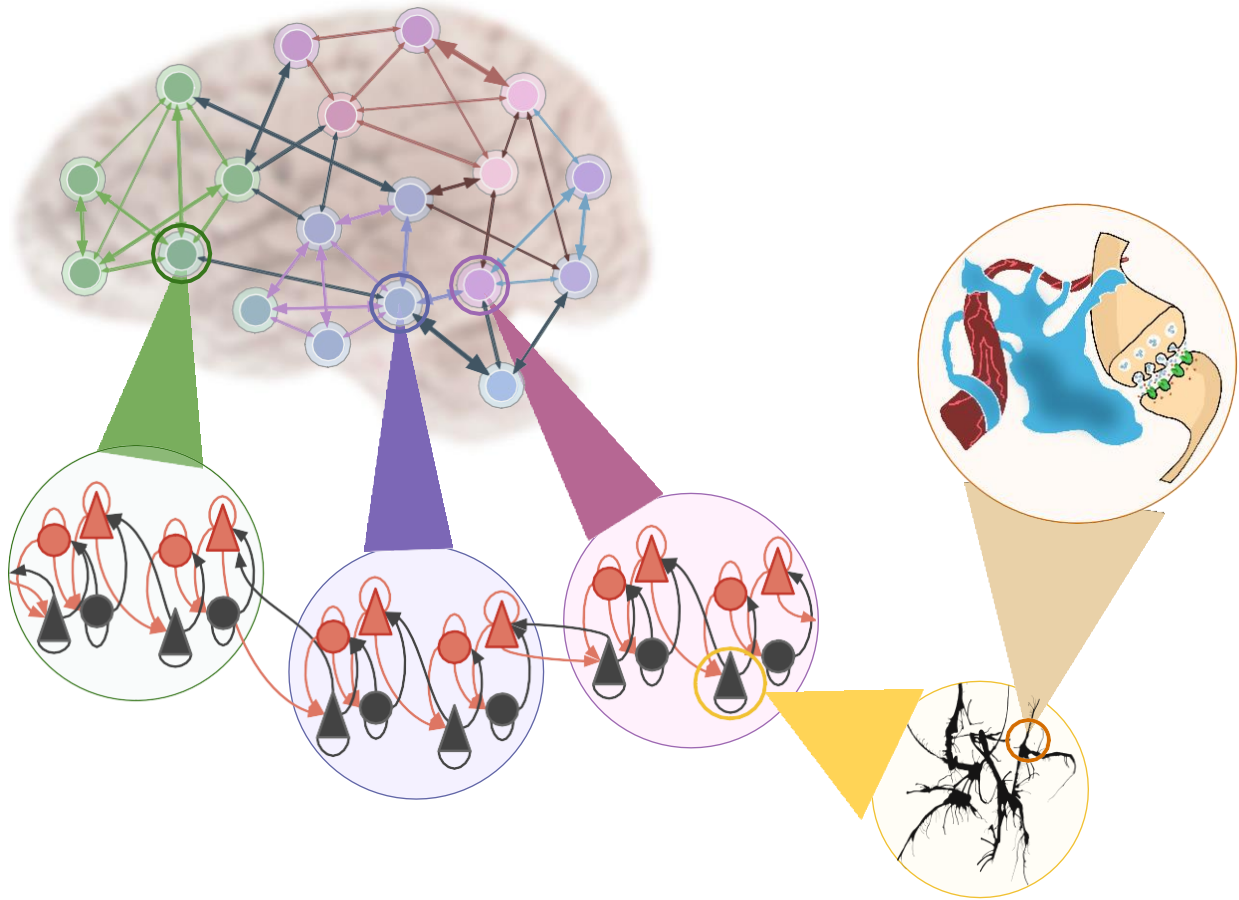

Figure 3: **Schematic graphic of nested hierarchies in the brain.** A representation of the repeating, nested Markov blanketed structure of the brain. Here, the same pattern of sparse coupling (and implicit conditional independencies)—i.e., the Markov blanket—repeats at every scale of interest; from neuronal compartments to the neural cell body, to microcircuits, brain regions, and whole-brain networks. Graphic inspired by Figure 1 in (Park & Friston, 2013).

These nested Markov blankets conform to the usual rules; namely, active states (i.e.,

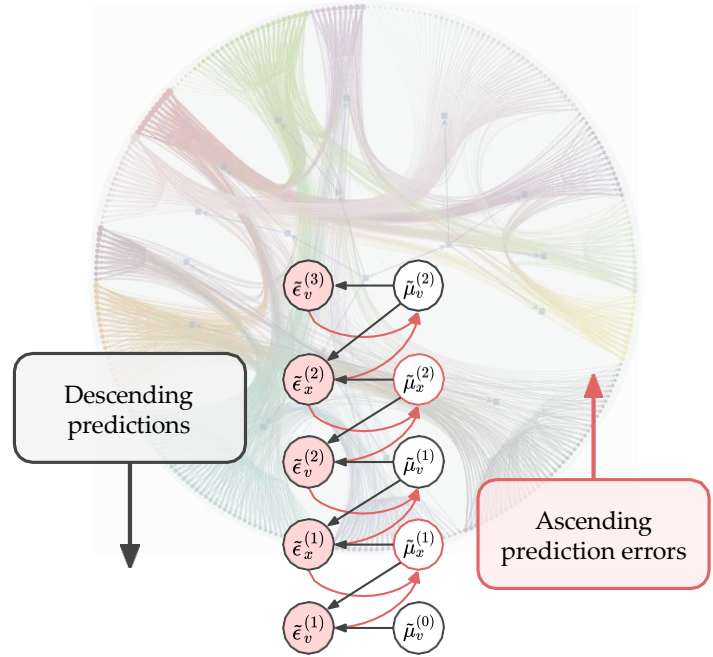

Figure 4: **Illustration of ascending and descending prediction errors in a brain.** Here, we represent a neural hierarchy implementing hierarchical predictive coding as a "vertical" stack, with "higher" levels providing context to "lower" sensory regions. This is equivalent to a representation of the same structure as a "centre-surround" architecture, with sensory levels at the periphery. Descending messages carry expectations and implement precision-weighting, while ascending messages carry precision-weighted prediction errors.

prediction errors) are not influenced by external states and sensory states (i.e., expectations) are not influenced by internal states. This follows from the rules of hierarchical predictive coding, in which prediction error units only influence expectation units and expectation units only influence error units (Friston, 2019b). In this example, the active states within the neuronal hierarchy exchange reciprocally with internal (sensory) states. Conversely, at the interface with the body, active states couple directly to neuromuscular junctions or secretory organs (Ondobaka, Kilner, & Friston, 2017). These correspond to the proprioceptive or autonomic (interoceptive) prediction errors that drive external states via classical reflexes (Adams, Shipp, & Friston, 2013; Seth & Friston, 2016).

Crucially, the depiction of hierarchical layers in terms of Markov blankets—and active and sensory states—has been equipped with covert action (Limanowski, 2017, 2022; Limanowski & Blankenburg, 2013; Safron, 2021a; Sandved-Smith et al., 2021); namely, the control of the precision of prediction error units at subordinate layers of the hierarchy by the expectations at superordinate levels. This is an example of a reciprocal or recurrent coupling between the sensory states of any Markov blanket and external states. This completes our pedagogical

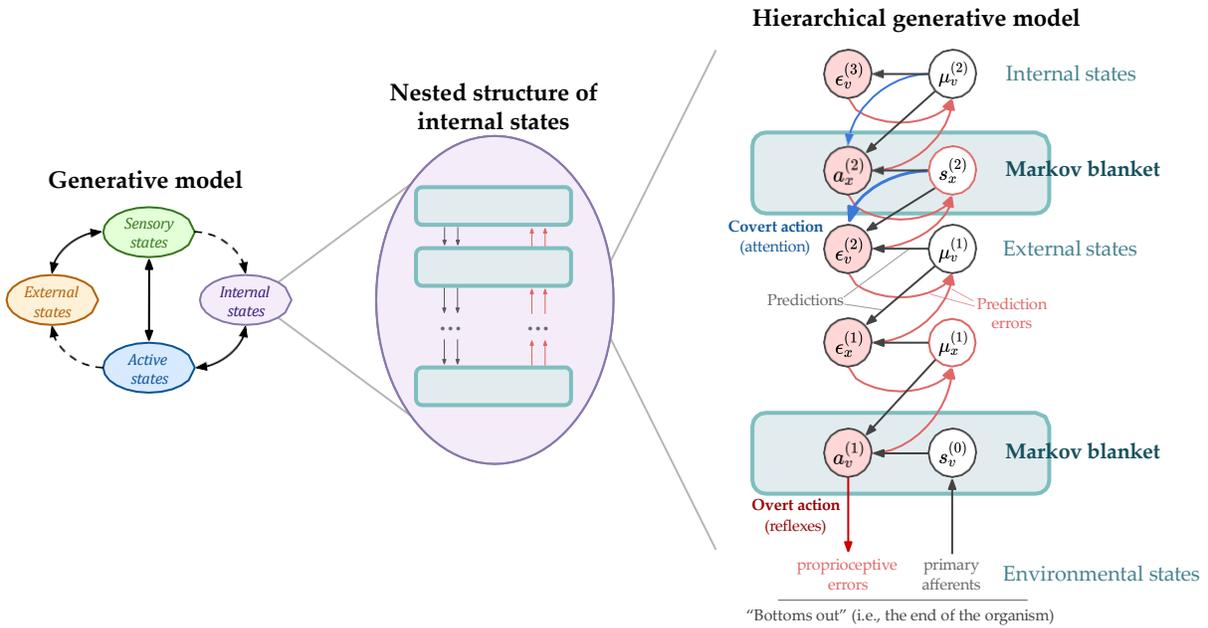

Figure 5: **Hierarchical predictive coding.** Left: Generative model. Middle: Abstracted depiction of *levels* of interacting Markov blankets that partition internal states. Right: This panel illustrates belief updating using a predictive coding scheme under a hierarchical (generative state space) model. This is a heuristic illustration of how this kind of message passing or belief propagation would work in the visual hierarchy. From a predictive coding perspective, the blue arrows correspond to the deployment of precision (i.e., attention) and the red arrows denote the influence of precision weighted prediction errors.

overview of the FEP formulation as it stands today in the predictive processing literature.

# 3 A FEP theoretic model for consciousness studies

We now consider our model for consciousness studies. This model is premised on the FEP, and follows from applying the constructs of nested Markov blankets and action selection in hierarchical generative models to explain well-established features of known neuroanatomy.

## 3.1 The brain as a hierarchy of nested blankets

As discussed above, in the FEP formulation, a Markov blanket is a mathematical structure that formalises the kind of separation and coupling that obtain between a thing and its environment. Crucially, all of the information available to an outside observer—about a given thing or system—is encoded on its Markov blanket (Fields & Marcianò, 2020; Friston, Fagerholm, et al., 2021). It should be noted, however, that the information encoded on the Markov blanket is not necessarily an exhaustive or complete description of the system, as the underlying states behind the blanket are inherently unknown—they are, by construction, unobservable and unmeasurable (Ramstead et al., 2021). This includes not only information about internal states enshrouded behind the Markov blanket itself, but also information about the nested Markov blankets within the system.

As discussed previously, every level of the neuronal hierarchy constitutes a processing stage that captures some relevant information about the environment and the self. The dynamical dependencies between these layers can be formalized as a Markov blanket between them. In the context of quantum information theory, the Markov blanket can be seen as separating processing stages (Fields & Glazebrook, 2022). This framing allows us to understand the flow of predictions and prediction errors as a process of reading and writing to successively nested holographic screens. The ascending prediction errors correspond to retrieving information from a given stage, which can be thought of as a measurement operator, or as reading from the screen. On the other hand, action on the environment corresponds to furnishing information to the environment, which can be thought of as a preparation operator, or writing to the screen. Only the information processed on the Markov blanket is classical, and can therefore constitute experience. The underlying processing may be (or may be modelled as) a form of quantum computation.

It should be noted that the information captured by each stage is not necessarily interpreted in the same way by different levels of the hierarchy. While, from our perspective, the information processed by lower levels can be interpreted as information about our world, it is not necessarily information about the world as perceived by the level or scale in question (Fields, Glazebrook, & Levin, 2022). The data to be processed consciously is distributed across the levels, with each successive level summarising everything that an observer would need to know about the levels below (and above) it, in order to understand the represented world at the corresponding level of abstraction.

This hierarchical organisation of the brain is thought to enable efficient processing of large amounts of information, and involves a process of coarse-graining, in which lower-level details are abstracted and integrated into higher-level summaries. As we ascend (and descend) its hierarchically nested layers, the information processed by each successive stage encodes increasingly long-timescale and global features of the environment (and the self as object). Representations towards the sensory levels of the hierarchy are domain-specific, while those that occupy deep positions within a centrifugal hierarchy are quintessentially amodal or multimodal (i.e., domain-general as opposed to domain-specific) and subsume interoception and exteroception (and proprioception). Thus, we can think of the set of nested Markov blankets as progressively coarse-graining the sensory information in the bottom-most or most peripheral sensory level of the brain (i.e., the information arriving at the sensory epithelia of the agent) in a kind of superposition of screens.

In summary, one can imagine centrifugal hierarchies of the sort depicted in Figure 5 as successively nested holographic screens; each reading and writing to each other at increasing levels of abstraction; i.e., coarse-graining over increasing temporal scales (Friston, Fagerholm, et al., 2021; Friston et al., 2017; George & Hawkins, 2009). The complete set of screens can thus be read as having a nested, holographic structure. Taking stock, already at this stage of construction, we are able to implement a version of the inner screen hypothesis (Fields,

Glazebrook, & Levin, 2021) by appealing only to the FEP to explain the known sparse anatomical structure of the brain.

## 3.2 Overt action, policy selection, and world-modelling

One deep question concerns how to pick out those screens whose content is, or maps onto, conscious experience. Clearly, most of the information that is encoded in the hierarchy of blankets does not correspond to anything that is experienced consciously. Which blanket(s) harness the information that maps onto the current contents of experience?

Following (Marchi & Hohwy, 2022) and (Nave et al., 2022), a core aspect of our proposal is that we can evaluate the capacity of each screen to act on its exterior to pick out the Markov blanket(s) that mediate the contents of experience. A formal or principled argument here is that if a system did not possess active states, there would be no manifestation of internal dynamics that called for an explanation; including the attributes of consciousness or experience. In other words, it would have no observable effect on the universe (other than the effect of sensory states that were caused by the universe). This epistemic argument suggests that no *detectably* conscious system can lack active states. More fundamentally, however, agents as such (of which conscious, human-like agents are a subset) are defined by their capacities both to act and to infer the consequences of action. But what kind of action are we talking about here?

The common thread—that underwrites the formulation above—is the reciprocal exchange across nested Markov blankets; or, equivalently, exchange of information via nested holographic screens (Fields & Levin, 2022). This existentially prerequisite sparse reciprocal coupling depends stipulatively upon *active states* that influence states exterior to the blanket, and consequent sensory impressions. This action can be overt (e.g., saccadic eye movements to sample salient sensory cues) or covert (e.g., selective attention or sensory attenuation mediated by neuromodulatory systems) (Ainley et al., 2016; Feldman & Friston, 2010; Hohwy, 2012; Pezzulo, 2012; Solms & Friston, 2018). Put simply, to consciously experience is to register change following an action. Covert action plays an asymmetrically important role in this account: while we can surely experience things over which we lack control, we cannot experience anything that it is impossible to attend to (a covert action).

Clearly—from the point of view of an inner screen—the action in question has to be covert (because it is hidden behind an outer screen or Markov blanket). This kind of action has been variously associated with mental action associated with attention: c.f., the premotor theory of attention (Rizzolatti et al., 1987). In terms of phenomenology, it has been proposed that mental action of this sort is a prerequisite for phenomenological opacity; namely, qualitative experience (Limanowski & Friston, 2018; Metzinger, 2003). From the perspective of the FEP, this kind of action is one in which active states *intervene causally* in external dynamics. Causal intervention has a straightforward and formal meaning here: it means that the causal coupling (i.e., Jacobian) between two external states depends upon the (active) intervention. Mathematically, this means that active states enter the equations of motion in a nonlinear fashion. (Note that causal intervention is a stronger requirement than simply saying external states depend upon active states, e.g., when active states enter the equations of motion

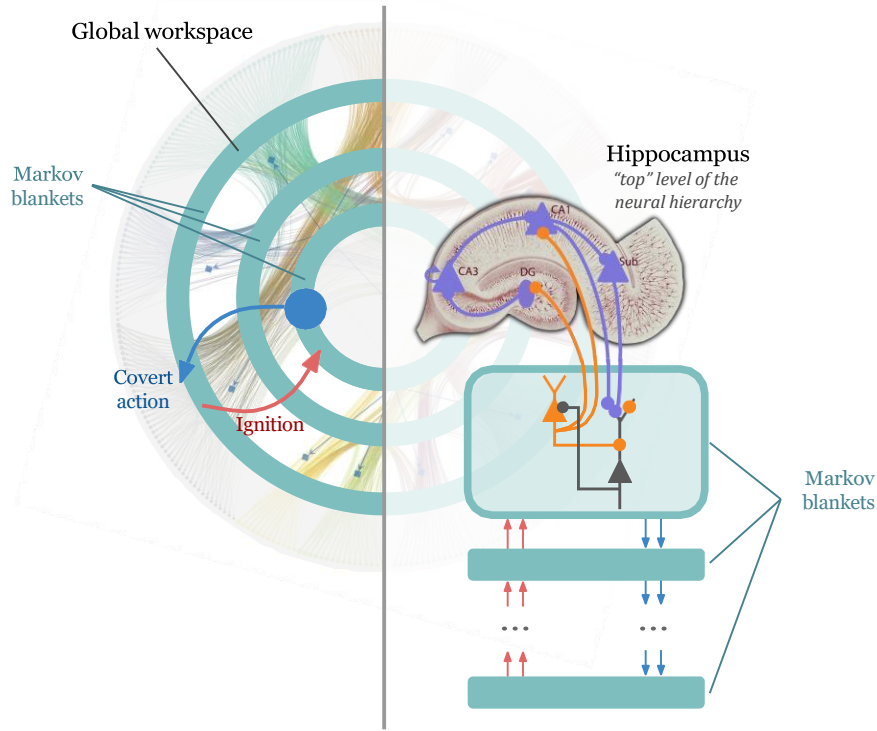

Figure 6: **Global workspace and nested Markov blankets.** Left: Schematic of nested levels of Markov blankets, highlighting message passing between blankets. Right: Superimposed illustration of the mechanisms playing out within levels of a neural hierarchy. In this example, we have placed the hippocampus at the apex of a centrifugal hierarchy. More generally, one would read the (irreducible) internal states of the deepest Markov blanket as homologous to a global (neuronal) workspace. The blue arrow on the left illustrates covert action; namely, engagement of neuromodulatory systems; such as the (subcortical) cells of origin of the ascending classical modulatory neurotransmitter systems (blue circle). In turn, the ensuing mental action (labelled "Ignition") modulates the message passing such that lower levels of the hierarchy gain access to the inner screen we have associated with the global workspace. Wiring diagram inspired and adapted from Fig. 3 in (Barron, Auksztulewicz, & Friston, 2020).

linearly).

When applied to neuronal systems, the implicit intervention can be read as neuromodulation; namely, changing the influence that one neuronal state has on another. Neuromodulation is generally taken to refer to a modulation of the excitability or gain of a neuron or neuronal population response to pre-synaptic inputs that is mediated by a change in post-synaptic conductances, as opposed to depolarisation. This is usually understood at the level of synaptic efficacy (Clark, 2013; Dayan, 2012). We now look more closely at what this kind of action would look like in applications of the FEP to hierarchical predictive processing.

## 3.3 Covert attention, precision-weighting, and self-modelling

A core aspect of our proposal concerns *covert* action. At first glance, the distinction between overt and covert actions may appear pronounced; however, from the perspective of Bayesian mechanics (Da Costa et al., 2021), they are just a statement that Markov blankets are constituted by active and sensory states. Indeed, from the point of view of certain neuronal populations, there may be no difference between the consequences of saccadic eye movements and a redeployment of spatial attention, as celebrated in the premotor theory of attention (Craighero & Rizzolatti, 2005; Rizzolatti & Craighero, 2010).

If we read covert action as selective attention (or sensory attenuation) and—in the setting of predictive coding—precision weighting of prediction errors, then internal (covert) action reduces to actively assigning precision to select certain sensory impressions on inner screens. Indeed, one can further argue that any qualitative experience necessitates a special kind of amodal representation (Barsalou, 1999; Palmeri & Gauthier, 2004) that can instantiate, for example, an attentional set. See (Sandved-Smith et al., 2021) for a worked example in computational phenomenology. In other words, representations of being in certain sentient states are necessary before one has a qualitative experience, as claimed in the higher-order thought theory of consciousness (Rosenthal, 2005). In particular, as active states can only be caused by internal states—and internal states represent something—then attention can only be deployed as the mental (covert) action that is conditioned on some internal state (of mind) that could be read as an attentional set or disposition.

On a more physiological note, it is telling that physiological fluctuations in conscious processing (i.e., different levels of arousal and sleep) are determined principally by ascending modulatory neurotransmitter systems (Hobson, 2009, 2021; Hobson & Friston, 2012; McCarley & Hobson, 1975). Crucially, exactly the same systems are targeted by pharmacological interventions in psychiatry and neurology (Anderson et al., 2013); namely, in altered states of consciousness encountered clinically. Finally, by definition, psychedelics act precisely (sic) at this level of sparse coupling among neuronal populations (Carhart-Harris & Friston, 2019; Muthukumaraswamy et al., 2013; Nour & Carhart-Harris, 2017; Safron, 2020b).

Figure 6 shows the same kind of architecture as in Figure 5, but now arranged to depict a global workspace as the interior of a deep Markov blanket. Access to the workspace rests upon *covert action*, i.e., *ignition* or attention that is mediated by the descending predictions of precision from the Markov blanket—and consequent ignition of ascending prediction errors that are written to the Markov blanket in proportion to their precision (Barron, Auksztulewicz, & Friston, 2020; Kanai et al., 2015; Rao & Ballard, 1999).

The formulation considered so far rests upon the active states of neuronal structures that influence their exterior to select or sample the right kind of sensory states. But does this mean that all Markov blankets are sentient? No. Only Markov blanketed systems that are endowed with precise or unambiguous dynamics will display this sort of behaviour. The argument is relatively straightforward: any thing of a sufficiently large scale or size will evince precise dynamics, because random fluctuations are averaged away. If we make the additional assumption that active states do not influence internal states, we can characterize the system as one that (looks as if it) plans its actions (Friston et al., 2022). In other words, the active states of certain things can be described as complying with two fundamental principles of Bayes optimality; namely, optimal Bayesian design (Lindley, 1956) and decision theory (Berger, 2011), respectively. Technically, the most likely paths of these kinds of things minimise *expected free energy*. Heuristically, they look as if they are planning and responding to salient cues to minimise uncertainty and render their belief updating as precise and unambiguous as possible. This, in turn, while perhaps not requiring nested inner screens as a matter of conceptual necessity, can as a matter of empirical fact only be realized via complex inner structure.

In short, some (distributed) neuronal structures—of a sufficiently large size—instantiate an attentional set in a way that rests upon internal beliefs about the most probable attentional set. It is, again, tempting to link this formulation with metacognitive and higher order thought formulations of sentient processing (Rosenthal, 2005); however, we will restrict ourselves to the following simple observation: To deploy an attentional set in a context-sensitive fashion, there must exist neuronal structures that entail generative models of attentional plans. If these are conditioned upon beliefs about context, then these beliefs are beliefs about one's attentional dispositions; namely, beliefs about perceptual set that underwrites qualitative experience.

Perhaps the simplest illustration of this argument is the difference between the perception of our heaving chest—during respiration—and our beating heart—during the cardiac cycle. Our respiration has a phenomenology. Our heartbeat does not (usually). The only difference between the processing of the respective interoceptive afferents is that we can consciously attend to the sensory consequences of breathing. If one subscribes to this account, awareness and phenomenology depend upon the ability to actively select the sensory evidence that underwrites belief updating. This is mental action: we are only aware of what we can control or select (in at least the minimal sense of choosing to attend to it). This is sometimes summarised as: "we cannot see unless we look, we cannot hear unless we listen, and we cannot feel unless we touch".

## 3.4 Consciousness and memory

We have argued that certain (deep) layers of the nested hologram—those that encode information that constitutes mental action—are uniquely associated with consciousness. There is a structural reason why it should be the case that the deep levels of the hierarchy are those associated with conscious experience.

The reason is fairly straightforward. If we think about how neuronal systems, each with their own blankets, are writing information to their local environments, it becomes clear that the top level in the hierarchy can only write downward—by definition. As such, the top level cannot store memories: namely, it cannot store classical information on its Markov blanket—except, that is, by acting on lower parts of the hierarchy. Similarly, the lower levels can only write up.

The hippocampus is often considered to be the top of the cortical hierarchy in memory prediction frameworks (Safron, Çatal, & Tim, 2022). The capacities of the hippocampus to orchestrate overall organismic and agentic trajectories in time-space could represent one of the primary determinants of the inherent thickness of the "specious present" (or "remembered present") and Husserlian accounts of time consciousness (Albarracin et al., 2022; Bergson, 2014). Accordingly, it has been suggested that the hippocampus acts as a source of "fictitious prediction-errors" and as the core locus of precision-weighting with respect to selection (Barron, Auksztulewicz, & Friston, 2020; Belluscio et al., 2012).

However, in terms of a control architecture, it may be more accurate to think of non-cortical systems as being hierarchically supraordinate, in the sense that it is these that mediate (mental or covert) action. These systems would include the homeostatic regulatory nuclei of the hypothalamus and septal area, and brainstem structures such as the superior and inferior colliculi, nucleus tractus solitarus and parabrachial nucleus, and periaqueductal grey (Damasio, 2018; Panksepp, 2004; Solms, 2013, 2021a; Swanson, 2000). Under the FEP, actions are specified by prior preferences for specific, phenotype-congruent sensory states. One could potentially go even further and extend ascription of control to the spinal reflex arcs, and even to the constraints and affordances of the body as it interacts with the world (Brooks, 1986; Paul, 2021; Tani, 2016), with its morphological intelligence constituting a kind of "prediction" or "embodied inference" (Friston, 2011), derived from phylogenetic learning (with phenotypes as generative models) and ontogenetic learning (e.g. experience-dependent, musculo-skeletal shaping) (Ramstead, Badcock, & Friston, 2018).

Structurally, the top level of any hierarchy cannot be the locus of externally stored memories, which are actions on the world that are then perceivable—like writing a sentence and then being able to read it or moving an object and then being able to see it. All of that is constructing memory by acting on its local world—or by acting on the Markov blanket, whichever way one wants to think of it. Crucially, the innermost screen or deepest Markov blanket in such a system is irreducible, in the sense that its internal states cannot be partitioned further. In other words, there are no Markov blankets in their interior. This speaks to a dense connectivity within certain (deep) neuronal systems or networks that—we suppose—are necessary for consciousness.

This foregrounds a very important feature of our model of consciousness: this "innermost screen" (which we have somewhat provocatively described as realizing a kind of "homunculus") must possess sufficient internal degrees of freedom, i.e. complexity, to act competently as a cybernetic controller, orchestrating the execution of actions on the world outside the overall organismic boundary via writing directly on the screen "below" it in the hierarchy. The consequences of these local actions are available to the high-level screen only via the mediation of sensory feedback routed through the intervening layers–rendering the latter functionally indispensable for the persistence of the higher-level boundary as such. The only way for the innermost level to induce memories about, or sense, the world is by acting upon subordinate levels. This causal mediation is a definitive aspect of agency and may be the definitive aspect of consciousness—as evinced through mental action. We have argued that this can be enabled only by the existence of a subsystem that is both irreducible (i.e. there exists no further Markov blanket partition of the internal states) and complex or expressive enough to act as a suitably flexible meta-controller (i.e., coordinating overt and covert action). Note that action *per se* is not in the province of internal states: active states are those that intervene on neural dynamics at subordinate levels. We have associated this causal intervention with neuromodulation.

Intriguingly, decomposition of the functional connectome suggests that the (intrinsic brain) networks associated with the transition between subliminal and conscious perception correspond to the "rich club" communication backbone of the brain (Lucini et al., 2019; Van Den Heuvel et al., 2012). These densely connected systems have descending connections to spinal motor pools, are associated with intentional control, and may constitute the most promising sites of intervention for brain-computer interfaces (Andersen, Aflalo, & Kellis, 2019; Jamieson, 2022; Kaas, Stepniewska, & Gharbawie, 2012).

It may seem counter-intuitive that the brainstem nuclei—the cells of origin of the ascending modulatory neurotransmitter systems—that manage arousal are construed as the active states of the innermost screen. However, this is entirely sensible, given their close association with levels of consciousness (and their regulatory role in altered states of consciousness, such as sleep). This embodies a strong hypothesis . Our claim is that all conscious systems must have an innermost screen that mediates covert action (e.g., arousal and attention). This means there is at least one irreducible set of internal states (i.e., bulk), behind the innermost screen, which determines the possible *contents* of consciousness. These could include objects, space-time, the self, explicit memories, plans, imaginative experience.

This suggests an evolutionary or phylogenetic picture in which "executive" neuronal structures—subcortical neuromodulatory systems—remain basically unchanged, while the bulk behind the innermost screen accumulates "mental solids," as (Solms, 2013) might call them. This accumulation may be reflected in cortical and subcortical systems that acquire progressively hierarchical (or heterarchical) structure, as we move from basal vertebrates to birds and mammals, or from basal molluscs to cephalopods. What is being added are, roughly speaking, concepts and processing power—or in the parlance of the quantum information theoretic formulation of the FEP, quantum reference frames (Fields et al., 2022). At first brush, this looks quite different from the usual triune picture, in which the cortex is associated with "higher thought" and thereby with the pinnacle of hierarchical organisation. However, the picture that follows from our model is not really that different: the cortex encodes the "higher concepts" that underwrite mental action at a subcortical level, while mental action underwrites the sensory impressions on the inner screen. Please see Box 1 for a summary of the ensuing model of consciousness.

# 4 Discussion

## 4.1 Theatres, attention, and uncertainty

Our discussion allows us to defuse some (but perhaps not all) tensions between accounts that emphasize the contributions of cortex to consciousness (such as IWMT; see (Safron, 2020a)) and those that emphasize the contributions of subcortical structures and in particular the brainstem (such as felt uncertainty theory; see (Solms, 2013, 2021b)); see, e.g., (Safron, 2021b; Solms, 2021a) for discussion. Indeed, the executive apex of hierarchical processing

> **Box 1: A minimal unifying model for consciousness under the FEP**
>
> Although the free energy principle (FEP) has little to say about consciousness *per se*, it offers a way of naturalising consciousness because the FEP—and its attendant Bayesian mechanics—share the same foundations with quantum, statistical and classical mechanics (Friston, 2019a; Sakthivadivel, 2022a). The implicit Bayesian mechanics inherits from something called a *particular partition* that individuates the internal states or *bulk* of a system from external states, via active and sensory states. This (Markov blanket) partition equips internal dynamics with an information geometry that can be read as *inference*; namely, representing—in a well-defined probabilistic sense—external dynamics (Ramstead et al., 2023). This information geometry is dual to the thermodynamics of the internal states *per se* (Friston, Wiese, & Hobson, 2020).
>
> Equipped with the notion of a Markov blanket, our model entails that consciousness is a property of a certain kind of Markov blanket, with the following properties. It is *irreducible* in the sense the internal states cannot be further partitioned at the scale in question. The *scale in question is classical* : i.e., a scale that admits classical mechanics and information (Fields, Glazebrook, & Levin, 2021). The internal states are influenced only by sensory states, while *active states intervene causally on external dynamics*; that is, change the causal coupling among external states, where causal is read in a control theoretic sense.
>
> **Neurobiology:** Applying this model to the brain means that consciousness is an attribute of brains that possess an irreducible Markov blanket, whose active states exert a *neuromodulatory* influence over external neuronal dynamics. For example, active states could constitute the cells of origin of the ascending modulatory neurotransmitter systems. The internal states would correspond to neuronal populations sending afferents to these cells (that are not reciprocated monosynaptically). These internal states could be read as a "dynamic core" or global neuronal workspace (Dehaene & Changeux, 2011); see our review below. In virtue of the Bayesian mechanics that describes these internal dynamics, they will necessarily evince a *generalised synchrony* with remaining brain (and body) states (Friston, Heins, et al., 2021).
>
> **Phenomenology:** Because the internal states cannot be further partitioned, no subset can be influenced by active states. This means internal states of an irreducible Markov blanket cannot be subject to *mental action*, in the sense of (Limanowski & Friston, 2018). In turn, this implies that any consciousness, should it exist, will be phenomenologically transparent. On this view, the only way that consciousness can be evinced is vicariously through active states. This means that consciousness is inherently *agential*. This aspect becomes formally more evident when one considers the FEP formulation of this (strange) kind of Markov blanket—that looks as if it has an intentional stance; in virtue of acting to *minimise expected free energy* (that has epistemic and instrumental components) (Friston et al., 2022).

corresponds not to the cortex—which controls subcortical processing—but rather, to a delicate array of structures in the upper brainstem—the structures that control the controller; e.g., neuromodulatory control of attentional processes: e.g., (Barron, Auksztulewicz, & Friston, 2020; Yu & Dayan, 2005). These structures lie at the centre of the brain. The only way that inner screens can change their sensory sectors is by acting upon lower levels, by emphasizing or de-emphasizing various conditional dependencies and implicit message passing, which we can construe as attentional selection or sensory attenuation. These processes are the province of the ascending neuromodulatory systems that we have associated with a sector of the inner screen; namely, the active states of an irreducible Markov blanket.

Our model of consciousness premised on the FEP might be read as a "minimal unifying model" of consciousness (Wiese, 2020). Such a MUM would conform to three criteria. First, a MUM must specify only necessary properties of (some forms of) conscious experience—which means that a MUM is minimal in that it does not aim (at least at first) to specify stronger, sufficient conditions for all conscious experience. Second, the model must include well defined, determinate descriptions of conscious experience that can be made more specific in further analyses, and that can be extended into more complex and focused models. Finally, a MUM is unifying, in that a MUM ought to integrate existing theoretical accounts of consciousness by foregrounding their common assumptions, which entails finding areas of overlap that correspond to necessary properties. When read as a MUM, the explanatory power of our model lies in its parsimony. We recognize, of course, that we have not established that the present model is a MUM; in particular, we have not shown that our model indeed integrates existing theoretical accounts of consciousness. We leave this task for future work.

Having said this, some novel predictions can be drawn from our model of consciousness. For example, if a holographic/quantum information theoretic formalism suggests that coherent locality of a "classical" variety is only achieved at an inner screen, then the identification of a locus for a kind of "Cartesian theatre," where cognitive re-presentations are formed—or alternatively, integrated and made available for widespread "broadcasting"—would indeed be both novel, and surprising to many (but not all) theorists.

We note that we have so far remained noncommittal about certain aspects of our proposal which may bear closer scrutiny from a philosophical point of view. In particular, an account like ours must squarely face the following argument: particular systems such as agents, according to the FEP, are individuated with respect to a Markov blanket that distinguishes them from their environments. Why then should we not ascribe consciousness to the "inner homunculus" individuated by the innermost screen in a hierarchical structure, rather than to the more extended system within which it is nested, which may (following Descartes, from whom we have taken some inspiration) in principle be replaced by another with the same functional profile, without (by hypothesis) affecting the contents of conscious experience?

This consideration foregrounds an important distinction among the commitments of our model. The presence of a Markov blanket possessing active states and housing a set of densely connected internal states at an appreciable scale appears to be the most obviously necessary constitutive condition for consciousness on our account, while the situation of this blanket within a set of concentrically nested screens may be read as a weaker or "merely causally necessary" enabling condition, as it is only in the context of such a nested holographic structure that such a complex but irreducible "inner bulk" could arise. This nested structure, we argue, is *de facto* essential to conscious systems in a physical universe anything like ours, as the dynamics of all relevant subsystems are coupled, even if sparsely. Moreover, the grounding of our argument in a relatively *a priori* scale-free account of the individuation of physical systems (i.e., the FEP) (Ramstead et al., 2023) confers a certain advantage over more traditional empirical theories in this context. Since we proceed from a principled and generic methodology for modelling dynamical systems, the minimal conditions on consciousness we describe are conceptually linked to the positing of a physical universe composed of Markov blankets at multiple scales.

# 5 Conclusion

The aim of this paper was to present a model of consciousness based directly on the variational free-energy principle (FEP). We first briefly introduced the FEP and unpacked its treatment of the way in which subsystems are individuated from and represent their surrounding environments, and then examined applications of the FEP to the nested and hierarchical neuroanatomy of the brain. We presented a putative model of consciousness based upon the FEP, focusing on the holographic structure of information processing in the brain, and how this structure engenders (overt and covert) action.

Swanson, L. W. (2000). Cerebral hemisphere regulation of motivated behavior. *Brain Research*, *886* (1-2), 113–164. https://doi.org/10.1016/s0006-8993(00)02905-x

Tani, J. (2016). *Exploring Robotic Minds: Actions, Symbols, and Consciousness as Self-Organizing Dynamic Phenomena*. Oxford University Press.

Tononi, G. (2015). Integrated information theory. *Scholarpedia*, *10* (1), 4164. https://doi.org/10.4249/scholarpedia.4164

Tononi, G., Boly, M., Massimini, M., & Koch, C. (2016). Integrated information theory: From consciousness to its physical substrate. *Nature Reviews Neuroscience*, *17* (7), 450–461. https://doi.org/10.1038/nrn.2016.44

Usher, M. (2001). A statistical referential theory of content: Using information theory to account for misrepresentation. *Mind and Language*, *16* (3), 331–334. https://doi.org/10.1111/1468-0017.00172

Van Den Heuvel, M. P., Kahn, R. S., Goñi, J., & Sporns, O. (2012). High-cost, high-capacity backbone for global brain communication. *Proceedings of the National Academy of Sciences*, *109* (28), 11372–11377. https://doi.org/10.1073/pnas.1203593109

Wiese, W. (2020). The science of consciousness does not need another theory, it needs a minimal unifying model. *Neuroscience of Consciousness*, *2020* (1), niaa013. https://doi.org/10.1093/nc/niaa013

Yu, A. J., & Dayan, P. (2005). Uncertainty, neuromodulation, and attention. *Neuron*, *46* (4), 681–692. https://doi.org/10.1016/j.neuron.2005.04.02634